\documentclass[preprint, 5p, twocolumn]{elsarticle}
\usepackage{amsmath,amssymb,bm,multirow,float,bbold}

\begin{document}
\begin{frontmatter}

\title{Abelian realization of phenomenological two-zero neutrino textures}

\author[a,b]{R. Gonz\'{a}lez Felipe}
\ead{ricardo.felipe@ist.utl.pt}

\author[rvt]{H. Ser\^{o}dio}
\ead{hugo.serodio@ific.uv.es}

\address[a]{Instituto Superior de Engenharia de Lisboa - ISEL, Rua
Conselheiro Em\'{\i}dio Navarro 1, 1959-007 Lisboa, Portugal}

\address[b]{Centro de F\'{\i}sica Te\'{o}rica de Part\'{\i}culas (CFTP), Instituto
Superior T\'{e}cnico, Universidade de Lisboa, Avenida Rovisco Pais 1, 1049-001
Lisboa, Portugal}

\address[rvt]{Departament de F\'{\i}sica Te\`{o}rica and IFIC, Universitat de
Val\`{e}ncia-CSIC, E-46100, Burjassot, Spain}


\begin{abstract}
In an attempt at explaining the observed neutrino mass-squared differences
and leptonic mixing, lepton mass matrices with zero textures have been
widely studied. In the weak basis where the charged lepton mass matrix is
diagonal, various neutrino mass matrices with two zeros have been shown to
be consistent with the current experimental data. Using the canonical and
Smith normal form methods, we construct the minimal Abelian symmetry
realizations of these phenomenological two-zero neutrino textures. The
implementation of these symmetries in the context of the seesaw mechanism
for Majorana neutrino masses is also discussed.
\end{abstract}

\end{frontmatter}

\section{Introduction}
\label{sec1}

In the last two decades, neutrino oscillation experiments have firmly
established the existence of neutrino masses and lepton mixing. However,
there remain several questions to be answered. From experiments we do not
know whether neutrinos are Dirac or Majorana particles, and whether CP is
violated or not in the lepton sector as it is for quarks (for recent reviews,
see e.g. Refs.~\cite{Strumia:2006db,Nunokawa:2007qh,Branco:2011zb}). In the
lack of a convincing theory to explain the origin of neutrino masses and
mixing, various approaches to the flavour puzzle have been pursued. In
particular, the imposition of some texture zeros in the neutrino mass matrix
allows to reduce the number of free parameters, and to establish certain
relations between the flavour mixing angles and mass ratios that could be
testable.

A common theoretical issue with mass (or coupling) matrices containing
vanishing entries is the origin of such texture zeros. It is possible to
enforce texture zeros in arbitrary entries of the fermion mass matrices by
means of discrete Abelian symmetries, e.g., the cyclic groups
$Z_n$~\cite{Grimus:2004hf,Grimus:2004az}. Yet the general methods commonly
used to obtain such patterns do not necessarily lead to their simplest
realization, i.e. with the smallest discrete Abelian group and number of
Higgs scalars.

In the basis where the charged lepton mass matrix is diagonal, the two-zero
textures for the effective neutrino mass matrix are phenomenological Ans\"{a}tze,
first studied by Frampton, Glashow and Marfatia (FGM) in
Ref.~\cite{Frampton:2002yf}. It turns out that only a subset of them is
presently compatible with the neutrino oscillation
data~\cite{Fritzsch:2011qv,Meloni:2014yea}. The aim of this work is to
construct the minimal Abelian symmetry realizations of these phenomenological
two-zero neutrino textures, and to study their implementation in extensions
of the standard model (SM) based on the seesaw mechanism for the neutrino
masses. In our search, we shall combine the canonical and Smith normal form
(SNF) methods, which have proved to be very successful in this
context~\cite{Petersen:2009ip,Ferreira:2010ir,Ivanov:2011ae,Serodio:2013gka,Ivanov:2013bka}.

\section{Two methods and their complementarity in model building}
\label{sec2}

There are two main approaches that can be used to study Abelian symmetries in
the Lagrangian: the canonical method  (see
Refs.~\cite{Ferreira:2010ir,Serodio:2013gka} and references therein) and the
SNF method~\cite{Ivanov:2011ae,Ivanov:2013bka}. In this section, we shall
briefly review these two methods and show their complementarity in studying
neutrino mass matrices with texture zeros.

In the canonical method, when dealing with Abelian symmetries, we represent
the generators of the symmetry group by diagonal phase matrices, i.e.
$\mathcal{S}=\text{diag}\,(e^{i\alpha},\,e^{i\beta},\cdots)$ for each set of
flavours. Consider, for example, the Yukawa-type interaction,
\begin{equation} \label{Yukint}
\overline{L}_\alpha Y^a_{\alpha\beta}R_\beta\,\Phi_a\,,
\end{equation}
with $\alpha,\beta=1,\cdots,n_f$ and $a=1,\cdots,n_h$; $L$, $R$, denote the
left-handed and right-handed fermion fields, respectively; $\Phi$ are the
Higgs fields. To study the symmetries of this interaction, we impose its
invariance under the field transformations
\begin{equation}
L\rightarrow \mathcal{S}_L L\,,\,R\rightarrow \mathcal{S}_R R\,,\,
\Phi \rightarrow \mathcal{S}_{\Phi}\Phi.
\end{equation}
This leads to the symmetry relations
\begin{equation} \label{symreq1}
\mathcal{S}_L^\dagger Y^a\mathcal{S}_R\, (\mathcal{S}_{\Phi})_a
=Y^a,
\end{equation}
where here, and henceforth, no summation over $a$ is assumed.

The above relations can be simplified by going to the Hermitian combinations
$H_L^a=Y^aY^{a\dagger}$ and $H_R^a=Y^{a\dagger}Y^a$, which have the symmetry
relations
\begin{equation} \label{symreq2}
\mathcal{S}_L^\dagger H_L^a\,\mathcal{S}_L=H_L^a\,,\quad
\mathcal{S}_R^\dagger H_R^a\,\mathcal{S}_R=H_R^a\,.
\end{equation}
From Eqs.~\eqref{symreq1} and~\eqref{symreq2}, one can then determine the
allowed textures of $Y^a$, for a given $n_f$ and independently of
$n_h$~\cite{Serodio:2013gka}. Note however that, in this approach, the
determination of the symmetry charges is a combinatorial problem which can be
very tedious.

In the Smith normal form method, a slightly different approach is followed.
The method uses the fact that if there are $n_F$ flavour fields (in our
example, $n_F=2n_f+n_h$), the Lagrangian would have an Abelian symmetry
$[U(1)]^{n_F}$ in the absence of phase sensitive terms. A term where a given
field appears only in combinations with its conjugated one, for example,
$\Phi_a^\dagger \Phi_a$, is not sensitive to the phase of that field. The
phase sensitive interactions will constrain the $U(1)$ groups by establishing
correlations between different groups or breaking them completely.

Let us take a simple example of two complex scalar fields $\phi_1$ and
$\phi_2$.  The phase insensitive Lagrangian is invariant under $U(1)\times
U(1)$, under which the fields transform as $\phi_1\rightarrow
(e^{i\alpha_1},1)\,\phi_1$ and $\phi_2\rightarrow (1,e^{i\alpha_2})\,\phi_2$.
In the presence of the interaction term $\phi_1^\dagger\phi_2$, we are no
longer able to rotate $\phi_1$ and $\phi_2$ arbitrarily. Both fields need to
be rotated with the same phase $\alpha$. Thus our initial symmetry is broken
down to a single continuous Abelian group, i.e. $U(1)\times U(1)\rightarrow
Z_1\times U(1)\,=\, U(1)$, with $\phi_{1,2} \rightarrow e^{i\alpha}
\phi_{1,2}$ as the symmetry transformation. If, on the other hand, the term
$\phi_1^2$ is added, the rephasing symmetry associated with $\phi_2$ will
remain unchanged, but the symmetry of $\phi_1$ is broken to $Z_2$. We then
get $U(1)\times U(1)\rightarrow Z_2\times U(1)$, with $\phi_{1} \rightarrow
(-1,1)\, \phi_{1}$ and $\phi_{2} \rightarrow (1,e^{i\alpha})\,\phi_{2}$ as
the symmetry transformation.

The idea of the SNF method is to deal with the symmetry breaking in a generic
way. To see how this is done we start by building a vector containing all the
fields. For the Yukawa interaction term of Eq.~\eqref{Yukint}, these are
\begin{equation}\label{vecF}
\left(\Phi_a\,,\,L_\alpha\,,\,R_\beta\right)\,.
\end{equation}
The sequence of the fields is irrelevant, but once an ordering is chosen it
must be kept until the end of the calculation. The steps for applying the
method are the following:

(i) For each phase sensitive interaction, we build a vector of the form of
Eq.~\eqref{vecF} where the entry $j$, associated with a particular flavour
field, is the number of fields minus the number of conjugated ones.

For example, considering the terms $\overline{L}_1R_2\Phi_1$ and
$\overline{L}_3R_2\Phi_2$, one writes
\begin{equation}
(\overbrace{1,0,\cdots}^{\Phi_a},\overbrace{-1,0,0,\cdots}^{L_\alpha},
\overbrace{0,1,\cdots}^{R_\beta})
\end{equation}
for the first term, and
\begin{equation}
\left(0,1,\cdots,0,0,-1,\cdots,0,1,\cdots\right)
\end{equation}
for the second one.

(ii) With the $k$ phase sensitive terms, we construct a $k \times n_F$ matrix
$D=\left\{d_{ij}\right\}$, where each row contains one of the vectors built
in (i). Since the Lagrangian must be invariant, the system of coupled
equations $d_{ij}\alpha_j=2\pi n_i$, with $n_i\in \mathbb{Z}$, has to be
satisfied.

(iii) We bring the matrix $D$ to its Smith normal form $D_{SNF}$, defined as
\begin{equation}\label{DSNF}
D_{SNF}=\text{diag}\,(d_1,\,d_2,\cdots,\,d_r,\,0,\cdots,\,0),
\end{equation}
with positive integers $d_i$ such that $d_i$ is a divisor of $d_{i+1}$ and
$r=\text{rank}(D)$. Note that the matrix $D_{SNF}$ is rectangular when $k
\neq n_F$, so that Eq.~\eqref{DSNF} means that everything else away from the
square block is also filled with zeros. For any integer value matrix $D$
there is a unique $D_{SNF}$ associated with it and related by
$D=R\,D_{SNF}\,C$. The matrices $R$ and  $C$ encode the operations (addition,
sign flip and permutation) on the rows and columns, respectively.

(iv) At this point, the system of equations has been transformed into a
system of  uncoupled equations $d_j\tilde{\alpha}_j=2\pi\tilde{n}_j$, with
$j=1,\cdots,r$, $\tilde{\alpha}_j=(C\alpha)_j$ and
$\tilde{n}_i=(R^{-1}n)_i$.\footnote{Note that, since the matrices $R$ and $C$
are invertible with $|\det| = 1$, we do not loose or generate any new
solution when switching from the system with $D$ to the system with
$D_{SNF}$. The symmetries are the same, and the solutions are in one-to-one
correspondence.} For $d_j\neq 0$ each equation corresponds to a $Z_{d_j}$
group, while for $d_j=0$ it corresponds to a $U(1)$ group. Thus the symmetry
of the Lagrangian has been broken down to
\begin{equation}\label{TotalSymm}
[U(1)]^{n_F} \rightarrow\; Z_{d_1}\times\cdots\times
Z_{d_r}\times [U(1)]^{n_F-r}\,.
\end{equation}
The original independent phases are now written as
\begin{equation}\label{phases}
\alpha_j=\left(\frac{2\pi}{d_1}(C^{-1})_{j1},\cdots,
\frac{2\pi}{d_r}(C^{-1})_{jr},\beta_{r+1},\cdots,\beta_{n_F}\right)\,.
\end{equation}

This simple procedure allows us to extract important information from the
presence of discrete and continuous symmetries in the Lagrangian. As
elegantly shown by Ivanov and Nishi~\cite{Ivanov:2013bka}, general conditions
for the possible model implementations without accidental $U(1)$'s can be
found. This approach has advantages over the canonical
one~\cite{Serodio:2013gka}, since the latter needs an explicit construction
to fully classify these models. Still, we remark that both methods have
interesting features that can be complementary from a model building
viewpoint. Note that, while in Eq.~\eqref{phases} the discrete phases are
predicted by the SNF method, the phases of the continuous groups are not.
Those can be easily obtained by the canonical method. It is also very common
in model building the implementation of specific zero-textures through the
use of symmetries. However, these zero-textures are in general present in the
mass terms and not in the interaction ones. To perform such a bottom-up
approach, it is useful to use the information on the allowed combinations of
textures, which can be easily constructed within the canonical method.

In this work, we shall use a bottom-up approach to answer the following
question:  Given a model where the neutrino mass matrix exhibits a two-zero
texture in the flavour basis, what are the minimal Abelian symmetries that
can be implemented to obtain such a pattern? In the next sections we shall
address this question using the canonical and SNF methods at different stages
of the problem.

\section{Textures from Abelian symmetries in the leptonic sector}
\label{sec3}

The origin of Majorana neutrino masses can be explained through the
introduction of the unique dimension-five operator compatible with the SM
gauge group. The leptonic interaction Lagrangian can be written as
\begin{align}\label{Lint}
-\mathcal{L}_{\text{int}}=\Pi^a_{\alpha\beta}\,\overline{\ell_{\alpha L}}\,
\phi_a\, e_{\beta R}+\frac{K_{\alpha\beta}^{ab}}{2\Lambda}\,
\left(\overline{\ell_{\alpha L}}\,\tilde{\phi}_a\right)
\left(\tilde{\phi}_b^T\,\ell_{\beta L}^c\right),
\end{align}
where $\Lambda$ is an effective energy scale, $\ell_L$ denotes the
left-handed lepton doublet fields and $e_R$ are the right-handed
charged-lepton singlets, $\tilde{\phi} = i \sigma_2 \phi^\ast$, and we allow
for the possibility of extra Higgs doublet fields $\phi_a$;
$a,b=1,\cdots,n_h$.

In the presence of flavour symmetries, two situations may occur: either the
high-dimensional operator is invariant under the full flavour symmetry or it
breaks the full flavour symmetry completely (or to a subgroup). The latter
case may arise from the ultraviolet (UV) completion of the model. The flavour
group at the UV level can be broken spontaneously through some additional
scalar fields (flavons), or it can even be broken explicitly by dimension
three (or less) operators. In what follows we shall focus only on the former
situation, i.e. when the dimension-five Weinberg operator is invariant under
the full flavour symmetry. In this case, a symmetry pattern can be defined
and the study of textures can be done in a model independent way.

Since we are only interested in the study of Abelian symmetries, following
the canonical approach we define the field transformations $\ell_{L}
\rightarrow \mathcal{S}_L\,\ell_L$, $e_{R}\rightarrow S_R\, e_R$, and $\Phi
\rightarrow \mathcal{S}_{\Phi}\, \Phi$, written in the basis where all the
transformations are diagonal unitary matrices, i.e.
\begin{align}
\mathcal{S}_{L} &= \text{diag}\left(e^{i\alpha_1}\,,e^{i\alpha_2},
\,e^{i\alpha_3}\right),\nonumber\\
\mathcal{S}_{R} &= \text{diag}\left(e^{i\beta_1}\,,e^{i\beta_2},
\,e^{i\beta_3}\right),\\
\mathcal{S}_{\Phi} &= \text{diag}\left(e^{i\theta_1}\,,e^{i\theta_2},
\,\cdots,\,e^{i\theta_N}\right)\nonumber.
\end{align}
These field transformations lead to the following symmetry relations
\begin{equation}
\mathcal{S}_L^\dagger\,\Pi^a\,\mathcal{S}_R\,e^{i\theta_a}=\Pi^a,\quad
\mathcal{S}_L^\dagger\,K^{ab}\,\mathcal{S}_L^\ast \,
e^{-i(\theta_a+\theta_b)}=K^{ab}.
\end{equation}
The textures for the Yukawa coupling matrices $\Pi^a$ are identical to the
quark  sector~\cite{Serodio:2013gka}, while for the matrix $K^{ab}$ we should
keep only the symmetric textures. Therefore, the allowed textures for
$K^{ab}$ are
\begin{equation}
P^T\left\{A_1\,,\, A_2\,,\,A_3\,,\,A_7\,,\,A_{12}
\right\} P
\end{equation}
for $P=\mathbb{1}$, $P_{13}$ or $P_{23}$, and
\begin{equation}
\left[P^\prime\left\{A_{13}\,,\,A_{15}\right\}P\right]_{\text{sym}}
\end{equation}
for any $P$ and $P^\prime$, with $P$ and $P^\prime$ denoting the $3 \times 3$
permutation matrices,
\begin{equation}
\begin{array}{ll}
&\;\;\;\mathbb{1}=\begin{pmatrix}
1&&\\
&1&\\
&&1
\end{pmatrix}\,,\quad
P_{12}=\begin{pmatrix}
&1&\\
1&&\\
&&1
\end{pmatrix},\\\\
&P_{13}=\begin{pmatrix}
&&1\\
&1&\\
1&&
\end{pmatrix}\,,\quad
P_{23}=\begin{pmatrix}
1&&\\
&&1\\
&1&
\end{pmatrix}\,,\\\\
&P_{123}=
\begin{pmatrix}
&&1\\
1&&\\
&1&
\end{pmatrix}\,, \quad P_{321}=
\begin{pmatrix}
&1&\\
&&1\\
1&&
\end{pmatrix}\,.
\end{array}
\end{equation}

The textures $A_i$ have the explicit patterns
\begin{align}\label{Textures}
\begin{split}
&A_1=
\begin{pmatrix}
\bm{\times}&\bm{\times}&\bm{\times}\\
\bm{\times}&\bm{\times}&\bm{\times}\\
\bm{\times}&\bm{\times}&\bm{\times}
\end{pmatrix}\,,\,
A_2=\begin{pmatrix}
\bm{\times}&\bm{\times}&\text{ }\\
\bm{\times}&\bm{\times}&\text{ }\\
\text{ }&\text{ }&\bm{\times}
\end{pmatrix}\,,\,\\
&A_3=\begin{pmatrix}
\text{ }&\text{ }&\bm{\times}\\
\text{ }&\text{ }&\bm{\times}\\
\bm{\times}&\bm{\times}&\text{ }
\end{pmatrix}\,,\,
A_7=
\begin{pmatrix}
\bm{\times}&\bm{\times}&\text{ }\\
\bm{\times}&\bm{\times}&\text{ }\\
\text{ }&\text{ }&\text{ }
\end{pmatrix}\,,
\\
&A_{12}=
\begin{pmatrix}
\text{ }&\text{ }&\text{ }\\
\text{ }&\text{ }&\text{ }\\
\text{ }&\text{ }&\bm{\times}
\end{pmatrix}\,,\,
A_{13}=
\begin{pmatrix}
\bm{\times}&\text{ }&\text{ }\\
\text{ }&\bm{\times}&\text{ }\\
\text{ }&\text{ }&\bm{\times}
\end{pmatrix}\,,\,\\
&A_{15}=
\begin{pmatrix}
\text{ }&\text{ }&\text{ }\\
\text{ }&\bm{\times}&\text{ }\\
\text{ }&\text{ }&\bm{\times}
\end{pmatrix},
\end{split}
\end{align}
where $\times$ denotes a nonzero matrix element. Since $\mathcal{S}_L$ acts
both on left and right of $K^{ab}$, we can group the possible textures into
three classes:
\begin{itemize}
\item[(1)] $\mathcal{S}_L=\mathbb{1}$ is completely degenerate. The only
    possible texture that can be implemented is $A_1$.

\item[(2)] $\mathcal{S}_L=P^T\text{diag}\left(1,1,e^{i\alpha}\right)P$,
    i.e. there is a two-fold degeneracy. This class can be seen as three
    subclasses, one for each matrix $P=\{\mathbb{1},\,P_{13},\,P_{23}\}$.
    The allowed textures are $P^T\left\{A_2,A_3,A_7,A_{12}\right\}P$.

\item[(3)]
    \mbox{$\mathcal{S}_L=\text{diag}\left(1,e^{i\alpha_1},e^{i\alpha_2}
    \right)$} is nondegenerate. For any $P$ and $P^\prime$, the allowed
    textures are  $ \left[P^\prime\left\{A_{12},A_{13},A_{15}\right\}
    P\right]_\text{sym}$.
    \end{itemize}

In the next section, we make use of the above classification in order to
reconstruct the FGM two-zero textures in terms of the textures allowed by the
Abelian symmetries.

\section{Decomposing the FGM two-zero textures}
\label{sec4}

The neutrino mass matrix is a symmetric matrix with six independent complex
entries. There are fifteen textures with two independent texture zeros,
usually classified into six categories ($\mathbf{A}_{1,2},
\mathbf{B}_{1,2,3,4}, \mathbf{C}, \mathbf{D}_{1,2}, \mathbf{E}_{1,2,3},
\mathbf{F}_{1,2,3}$)~\cite{Frampton:2002yf}. While this classification may be
advantageous for phenomenological studies, from the symmetry viewpoint it is
more convenient to group them in a slightly different
way~\cite{Cebola:2013hta}:
\begin{equation} \label{2nutexturesII}
\begin{array}{lll}
\widetilde{\mathbf{A}}&\equiv&\left\{\mathbf{A}_1,\,\mathbf{A}_2=P_{23}
\mathbf{A}_1 P_{23},\,\mathbf{B}_3=P_{12} \mathbf{A}_1 P_{12},\,\right.\\
&&\,\mathbf{B}_4=P_{321} \mathbf{A}_1 P_{123},\,\mathbf{D}_1=P_{123}
\mathbf{A}_1 P_{321},\\
&&\left.\,\mathbf{D}_2=P_{13} \mathbf{A}_1 P_{13}\right\},\\\\
\widetilde{\mathbf{B}}&\equiv&\left\{\mathbf{B}_1,\,\mathbf{B}_2=P_{23}
\mathbf{B}_1 P_{23},\,\mathbf{E}_3=P_{12} \mathbf{B}_1 P_{12}\right\},\\\\
\widetilde{\mathbf{C}}&\equiv&\left\{\mathbf{C},\,\mathbf{E}_1=P_{321}
\mathbf{C} P_{123},\,\mathbf{E}_2=P_{12} \mathbf{C} P_{12}\right\},\\\\
\widetilde{\mathbf{F}}&\equiv&\left\{\mathbf{F}_1,\,\mathbf{F}_2=P_{12}
\mathbf{F}_1 P_{12},\,\mathbf{F}_3=P_{13} \mathbf{F}_1 P_{13}\right\},
\end{array}
\end{equation}
where
\begin{equation}
\begin{array}{l}
\mathbf{A}_1:\,
\begin{pmatrix}
0&0&\times\\
0&\times&\times\\
\times&\times&\times
\end{pmatrix}\,,\quad
\mathbf{B}_1:\,
\begin{pmatrix}
\times&\times&0\\
\times&0&\times\\
0&\times&\times
\end{pmatrix}\,,\\\\
\;\;\mathbf{C}:\,
\begin{pmatrix}
\times&\times&\times\\
\times&0&\times\\
\times&\times&0
\end{pmatrix}\,,\quad
\mathbf{F}_1:\,
\begin{pmatrix}
\times&0&0\\
0&\times&\times\\
0&\times&\times
\end{pmatrix}\,.
\end{array}
\end{equation}

In the flavour basis, the only two-zero neutrino textures that can be
obtained with a single scalar Higgs doublet are the ones in class
$\widetilde{\mathbf{F}}$. This is a direct consequence of
Eqs.~\eqref{Textures}, since no other two-zero texture is in the set of
allowed textures. This class is however phenomenologically excluded.
Therefore, in order to implement the phenomenologically viable two-zero
textures $\mathbf{A}_i,\, \mathbf{B}_i$ and $\mathbf{C}$, an extended scalar
sector with $n_h \geq 2$ is needed. Following a path similar to that of
Ref.~\cite{Serodio:2013gka}, we shall use a bottom-up approach in order to
find the possible ways of implementing the above two-zero textures.

To illustrate our approach, we consider class $\widetilde{\mathbf{B}}$ and
choose the texture $\mathbf{B}_1$. The first step is to decompose the
zero-texture $\mathbf{B}_1$ in the largest possible set of textures, keeping
in mind that the effective neutrino mass matrix should be symmetric. We
obtain the decomposition
\begin{align}
\begin{split}
\mathbf{B}_1&=\footnotesize{\begin{pmatrix}
&\times&\\
\times&&\\
&&
\end{pmatrix}+
\begin{pmatrix}
&&\\
&&\times\\
&\times&
\end{pmatrix}+
\begin{pmatrix}
\times&&\\
&&\\
&&
\end{pmatrix}+
\begin{pmatrix}
&&\\
&&\\
&&\times
\end{pmatrix}}\\ \\
&=P_{321}A_{15}P_{13}\oplus
A_{15}P_{23}\oplus P_{13}A_{12}P_{13}\oplus A_{12}.
\end{split}
\end{align}
The maximal number of different textures that build $\mathbf{B}_1$ is four.
Any model with more than four interaction terms will have necessarily
repeated textures.

The next step is to reduce the above decomposition from four to three
textures. There are six possibilities. Summing the first and second textures
we get the decomposition
\begin{equation}
\begin{split}
\mathbf{B}_1&=\begin{pmatrix}
&\times&\\
\times&&\times\\
&\times&
\end{pmatrix}+
\begin{pmatrix}
\times&&\\
&&\\
&&
\end{pmatrix}+
\begin{pmatrix}
&&\\
&&\\
&&\times
\end{pmatrix}\\ \\
&=P_{23}A_{3}P_{23}\oplus P_{13}A_{12}P_{13}\oplus A_{12}.
\end{split}
\end{equation}
The first texture is not compatible with the other two. Indeed, although the
three textures belong to class (2), the first one belongs to the subclass
with $P=P_{23}$, while the second and third textures are in the subclasses
with $P=P_{13}$ and $P=\mathbb{1}$, respectively.

Summing the first and third textures we get the decomposition
\begin{equation}
\mathbf{B}_1=\begin{pmatrix}
\times&\times&\\
\times&&\\
&&
\end{pmatrix}+
\begin{pmatrix}
&&\\
&&\times\\
&\times&
\end{pmatrix}+
\begin{pmatrix}
&&\\
&&\\
&&\times
\end{pmatrix}.
\end{equation}
In this case, the first texture does not exist in the set of textures allowed
by the Abelian symmetries.

Summing the first and fourth textures we obtain
\begin{equation}
\begin{split}
\mathbf{B}_1&=\begin{pmatrix}
&\times&\\
\times&&\\
&&\times
\end{pmatrix}+
\begin{pmatrix}
&&\\
&&\times\\
&\times&
\end{pmatrix}+
\begin{pmatrix}
\times&&\\
&&\\
&&
\end{pmatrix}\\ \\
&=A_{13}P_{12}\oplus A_{15}P_{23}\oplus P_{13}A_{12}P_{13},
\end{split}
\end{equation}
which are compatible textures belonging to class (3). We can continue the
above procedure, obtaining at the end two more implementations with three
textures, namely, $A_{13}P_{23}\oplus P_{321}A_{15}P_{13}\oplus A_{12}$, and
$P_{321}A_{15}P_{13}\oplus A_{15}P_{23}\oplus P_{12}A_{15}P_{12}$.

Similarly, matrices with only two textures can be constructed.  We need to
look just at the allowed cases with three textures. It turns out that only in
the first case there exists a viable implementation of two textures, which is
given by the decomposition $A_{13}P_{12}\oplus A_{13}P_{23}$.

Below we summarize the allowed texture decompositions for the three classes
$\widetilde{\mathbf{A}}$, $\widetilde{\mathbf{B}}$ and
$\widetilde{\mathbf{C}}$:
\begin{align} \label{decompclassA}
\begin{array}{ll}
\widetilde{\mathbf{A}}&\text{Class}\\\\
\text{(a)}&P_{12}A_{15}P_{321}\oplus A_{15}P_{23}\oplus P_{23}A_{12}P_{23}
\oplus A_{12},\\\\
\text{(b)}&A_{13}P_{13}\oplus A_{15}P_{23}\oplus A_{12},\\\\
\text{(c)}&P_{12}A_{15}P_{321}\oplus A_{15}P_{23}\oplus A_{15}.\\
\end{array}
\end{align}
\begin{align} \label{decompclassB}
\begin{array}{ll}
\widetilde{\mathbf{B}}&\text{Class}\\\\
\text{(a)}&P_{321}A_{15}P_{13}\oplus A_{15}P_{23}\oplus P_{13}A_{12}P_{13}
\oplus A_{12},\\\\
\text{(b)}&A_{13}P_{12}\oplus A_{15}P_{23}\oplus P_{13}A_{12}P_{13},\\\\
\text{(c)}&P_{321}A_{15}P_{13}\oplus A_{15}P_{23}\oplus P_{12}A_{15}P_{12},\\\\
\text{(d)}&A_{13}P_{12}\oplus A_{13}P_{23}.\\
\end{array}
\end{align}
\begin{align} \label{decompclassC}
\begin{array}{ll}
\widetilde{\mathbf{C}}&\text{Class}\\\\
\text{(a)}&P_{321}A_{15}P_{13}\oplus P_{123}A_{15}P_{12}\oplus A_{15}P_{23}\oplus
P_{13}A_{12}P_{13},\\\\
\text{(b)}&A_{13}P_{23}\oplus P_{321}A_{15}P_{13}\oplus P_{123}A_{15}P_{12}.
\end{array}
\end{align}

Next we shall study the realization of the above decompositions through the
introduction of Abelian symmetries in the context of 2HDM.

\section{Implementation of FGM two-zero textures in 2HDM}
\label{sec5}

In our previous analysis, we have made use of the canonical method to obtain
the possible textures that implement the FGM two-zero textures. In this
section, we shall use the SNF method in order to find the corresponding
Abelian symmetries. The vector containing the relevant fields of the leptonic
sector, in the presence of the $d=5$ effective operator, is defined as
$\left(\phi_a,\,\ell_{\alpha L},\,e_{\beta R}\right)$.

The FGM two-zero textures are written in the basis where the charged leptons
are diagonal. Inserting this requirement into the matrix $D$ and performing
some simple operations on rows and columns (see steps (ii) and (iii) for the
SNF method described in section~\ref{sec2}), it will lead to a structure of
the form
\begin{equation}\label{Dneutrinos}
D_{k\times (n_h+6)}\sim\left(
\begin{array}{c|c}
\mathbb{0}_{3\times (n_h+3)}&\mathbb{1}_{3}\\
\hline
\mathcal{D}_{(k-3)\times (n_h+3)}&\mathbb{0}_{(k-3)\times 3}
\end{array}
\right).
\end{equation}
This implies that, in order to find the symmetries in the leptonic sector, we
only need to look at the matrix $\mathcal{D}$ and the associated field vector
$(\phi_a,\ell_\alpha)$ of the effective operator
interactions.\footnote{Notice that there are also Abelian symmetries in the
charged lepton sector. For example, if only one Higgs doublet is coupled to
charged leptons, then $n_F=7$ and $r=3$ (three nonzero interactions). This
implies the presence of four Abelian continuous symmetries: the global
hypercharge and $U(1)_e\times U(1)_\mu\times U(1)_\tau$. Nevertheless, the
neutrino interaction terms break the latter symmetry down, so that the
flavour symmetry in the whole Lagrangian is dictated by the neutrino
textures.}

First we note that there are always $n_h+3-r$ continuous Abelian symmetries
in our subsystem, as can be seen from Eq.~\eqref{TotalSymm}. The rank of
$\mathcal{D}$ is at most $4$, which is the number of distinct interactions in
the effective term. Therefore, for the FGM two-zero textures there exist
$n_h-1$ continuous Abelian symmetries. Since the global hypercharge $U(1)_Y$
is always present, only two Higgs doublet models (2HDM) may avoid additional
continuous symmetries in the leptonic sector. The most dangerous continuous
symmetries are the ones present in the scalar potential, since they may lead
to (pseudo-) Goldstone bosons. For now, we shall only focus on the leptonic
sector in order to find the minimal discrete Abelian realizations of the FGM
two-zero textures in 2HDM. We shall then comment on the symmetries of the
scalar sector at the end of this section.

\begin{table*}[t]
\caption{\label{TabSym} Two-zero texture decompositions, Higgs combinations and
the corresponding Abelian symmetries for classes $\widetilde{\mathbf{A}}$,
$\widetilde{\mathbf{B}}$ and $\widetilde{\mathbf{C}}$.}
\begin{center}
\begin{tabular}{cccc}
\hline
Texture&Texture decomposition&Higgs combination&Symmetry\\
\hline
\multirow{6}{*}{
$\widetilde{\mathbf{A}}=
\begin{pmatrix}
0&0&\times\\
0&\times&\times\\
\times&\times&\times
\end{pmatrix}
$
}&&&\\
 &\multirow{3}{*}{$\begin{array}{c}
 A_{13}P_{13}\oplus A_{15}P_{23}\oplus A_{12}\\[7pt]
 {\footnotesize\begin{pmatrix}
 &&\times\\
 &\times&\\
 \times&&
 \end{pmatrix}+\begin{pmatrix}
 &&\\
 &&\times\\
 &\times&
 \end{pmatrix}+\begin{pmatrix}
 &&\\
 &&\\
 &&\times
 \end{pmatrix}}
 \end{array}$}& \multirow{3}{*}{$[(1,1),(1,2),(2,2)]$}&
 \multirow{3}{*}{$Z_2\times U(1)$}\\[7pt]
&&&\\
&&&\\
&&&\\
\hline
&&&\\
\multirow{14}{*}{
$\widetilde{\mathbf{B}}=
\begin{pmatrix}
\times&\times&0\\
\times&0&\times\\
0&\times&\times
\end{pmatrix}
$
} &\multirow{3}{*}{$\begin{array}{c}
 A_{13}P_{12}\oplus A_{15}P_{23}\oplus P_{13}A_{12}P_{13}\\[7pt]
 {\footnotesize\begin{pmatrix}
 &\times&\\
 \times&&\\
 &&\times
 \end{pmatrix}+\begin{pmatrix}
 &&\\
 &&\times\\
 &\times&
 \end{pmatrix}+\begin{pmatrix}
 \times&&\\
 &&\\
 &&
 \end{pmatrix}}
 \end{array}$}&\multirow{4}{*}{$
 [(1,1),(2,2),(1,2)]
 $}&\multirow{4}{*}{$
 Z_2\times Z_5
 $}\\
&&&\\
&&&\\
&&&\\
&&&\\
& \multirow{3}{*}{$\begin{array}{c}
P_{321}A_{15}P_{13}\oplus A_{15}P_{23}\oplus P_{12}A_{15}P_{12}\\[7pt]
 {\footnotesize\begin{pmatrix}
 &\times&\\
 \times&&\\
 &&
 \end{pmatrix}+\begin{pmatrix}
 &&\\
 &&\times\\
 &\times&
 \end{pmatrix}+\begin{pmatrix}
 \times&&\\
 &&\\
 &&\times
 \end{pmatrix}}
 \end{array}$}&\multirow{3}{*}{$[(1,1),(2,2),(1,2)]$}&\multirow{3}{*}{$Z_8$}\\
&&&\\
&&&\\
&&&\\
&&&\\
& \multirow{3}{*}{$\begin{array}{c}
A_{13}P_{12}\oplus A_{13}P_{23}\\[7pt]
 {\footnotesize\begin{pmatrix}
 &\times&\\
 \times&&\\
 &&\times
 \end{pmatrix}+\begin{pmatrix}
 \times&&\\
 &&\times\\
 &\times&
 \end{pmatrix}}
 \end{array}$}&\multirow{3}{*}{$[(1,1),(2,2)]$}&\multirow{3}{*}{$
 Z_2\times Z_2\times Z_3 $}\\
&&&\\
&&&\\
&&&\\[7pt]
\hline
&&&\\
\multirow{3}{*}{
$\widetilde{\mathbf{C}}=
\begin{pmatrix}
\times&\times&\times\\
\times&0&\times\\
\times&\times&0
\end{pmatrix}
$
} & \multirow{3}{*}{$\begin{array}{c}
 A_{13}P_{23}\oplus P_{321}A_{15}P_{13}\oplus P_{123}A_{15}P_{12}\\[7pt]
 {\footnotesize\begin{pmatrix}
 \times&&\\
 &&\times\\
 &\times&
 \end{pmatrix}+\begin{pmatrix}
 &\times&\\
 \times&&\\
 &&
 \end{pmatrix}+\begin{pmatrix}
 &&\times\\
 &&\\
 \times&&
 \end{pmatrix}}
 \end{array}$}&\multirow{3}{*}{$ \left[(1,2),(1,1),(2,2)\right]$}&
 \multirow{3}{*}{$U(1)$}\\
&&&\\
&&&\\
&&&\\[9pt]
\hline
\end{tabular}
\end{center}
\end{table*}

Working in a framework with two scalar fields, $\phi_{1,2}$, we can only form
three distinct combinations: $(\phi_1)^2$, $(\phi_2)^2$ and $\phi_1\phi_2$.
This implies that the three cases labelled (a) in
Eqs.~\eqref{decompclassA}-\eqref{decompclassC} are automatically excluded. To
see how the symmetries can be straightforwardly determined using the SNF
method, let us consider, for instance, the decomposition (b) in class
$\widetilde{\mathbf{B}}$. One possibility for the interaction Lagrangian is
\begin{align}
\begin{split}
-\mathcal{L}_{int}^{d=5}=\; & \overline{\ell_L}\left[
A_{13}P_{12}(\phi_1^\ast)^2+A_{15}P_{23} (\phi_2^\ast)^2\right.\\
&\left.
\quad+P_{13}A_{12}P_{13}(\phi_1^\ast\phi_2^\ast)
\right]\ell_L^c\,.
\end{split}
\end{align}

From this Lagrangian we build the matrix $\mathcal{D}$,
\begin{equation}
\mathcal{D}=\begin{pmatrix}
-2&0&-1&-1&0\\
-2&0&0&0&-2\\
0&-2&0&-1&-1\\
-1&-1&-2&0&0
\end{pmatrix}\,,
\end{equation}
which has the Smith normal form $\mathcal{D}_{SNF}=\text{diag}(1,1,1,10)$,
and leads to the symmetry $Z_{10}\sim Z_2\times Z_5$. As explained in step
(iv) of section~\ref{sec2}, one can extract the discrete charges for the
flavour symmetry using the information coming from the operations on columns
(i.e. from matrix $C$). We get
\begin{equation}
Z_2\times Z_5:\,\left\{
\begin{array}{l}
\phi_1\rightarrow (-1,1)\,\phi_1\,,\quad \phi_2 \rightarrow (-1,\omega_5^2)\,\phi_2\,,\\
\ell_{eL} \rightarrow (1,\omega_5^4)\, \ell_{e L}\,,\quad
\ell_{\mu L} \rightarrow (1,\omega_5)\,\ell_{\mu L}\,,\\
\ell_{\tau L} \rightarrow (1,1)\,\ell_{\tau L}\,,
\end{array}\right.
\end{equation}
where $\omega_n = e^{i 2\pi/n}$.

Note that the $Z_2$ group is simply the discrete lepton number that remains
after the explicit breaking of $U(1)_L$ by the $d=5$ effective operator.
Since in this implementation we have coupled $(\phi^\ast_1)^2$ to the first
texture, $(\phi_2^\ast)^2$ to the second one, and $\phi_1^\ast\phi_2^\ast$ to
the third one, we denote this Higgs combination by $[(1,1),(2,2),(1,2)]$.
Checking all the other Higgs combinations, we conclude that none of them has
a symmetry implementation.

The same procedure can be used to find the symmetry implementations for all
classes. In Table~\ref{TabSym} we summarize the texture decompositions that
can be implemented, their Higgs combinations and the associated Abelian
symmetries. For completeness, we present below the symmetry charges for the
other cases given in Table~\ref{TabSym}. For class $\widetilde{\mathbf{A}}$
we have
\begin{equation}\label{ClassA}
\begin{array}{ll}
Z_2\times U(1): & Z_2\times Z_5: \\\\
\left\{
\begin{array}{l}
\phi_1\rightarrow (-1,1)\,\phi_1\,,\\
\phi_2\rightarrow (-1,e^{i\alpha})\,\phi_2\,,\\
\ell_{eL}\rightarrow (1,e^{i\alpha})\,\ell_{e L}\,,\\
 \ell_{\mu L}\rightarrow (1,1)\,\ell_{\mu L}\,,\\
  \ell_{\tau L}\rightarrow (1,e^{-i\alpha})\,\ell_{\tau L}\,,
\end{array}\right.\supset
&
\left\{
\begin{array}{l}
\phi_1\rightarrow (-1,1)\,\phi_1\,,\\
\phi_2\rightarrow (-1,\omega_5)\,\phi_2\,,\\
\ell_{eL}\rightarrow (1,\omega_5)\,\ell_{e L}\,,\\
 \ell_{\mu L}\rightarrow (1,1)\,\ell_{\mu L}\,,\\
  \ell_{\tau L}\rightarrow (1,\omega_5^4)\,\ell_{\tau L}\,.
\end{array}\right.
\end{array}
\end{equation}
While the SNF method points in this case to the existence of a continuous
Abelian symmetry $U(1)$, there is a minimal discrete Abelian realization
$Z_5$ that leads to this texture. This symmetry can be found within the
canonical method (see Table~\ref{tablechains} in~\ref{app1}).

For the remaining decompositions of class $\widetilde{\mathbf{B}}$ we get
\begin{equation}
Z_8:\,\left\{
\begin{array}{l}
\phi_1\rightarrow \phi_1\,,\quad\phi_2\rightarrow \omega_8^2\,\phi_2\,,\\
\ell_{eL}\rightarrow \omega^3_8\,\ell_{e L}\,,\quad \ell_{\mu L}\rightarrow
\omega_8^5\,\ell_{\mu L}\,,\\ \ell_{\tau L}\rightarrow \omega^7_8\,\ell_{\tau L}\,,
\end{array}\right.
\end{equation}
and
\begin{equation}
Z_2\times Z_2\times Z_3:\left\{
\begin{array}{l}
\phi_1\rightarrow (1,1,1)\,\phi_1\,,\\
\phi_2\rightarrow (1,-1,\omega_3)\,\phi_2\,,\\
\ell_{eL}\rightarrow (-1,-1,\omega_3^2)\,\ell_{e L}\,,\\
\ell_{\mu L}\rightarrow (-1,-1,\omega_3)\,\ell_{\mu L}\,,\\
 \ell_{\tau L}\rightarrow (-1,-1,1)\,\ell_{\tau L}\,,
\end{array}\right.
\end{equation}
where the group $Z_6$, coming from the Smith normal form, has been decomposed
as $Z_2\times Z_3$.

Finally, for class $\widetilde{\mathbf{C}}$ we obtain\footnote{The Higgs
field transformation for the discrete symmetry $Z_8$ is brought to the form
given in Eq.~\eqref{CaseC} by means of the global phase transformation
$\phi_a\rightarrow \omega_8\,\phi_a$ and $\ell_{\alpha L}\rightarrow
\omega_8^{-1}\,\ell_{\alpha L}$.}
\begin{equation}\label{CaseC}
\begin{array}{ll}
U(1): &Z_8:\\\\
\left\{
\begin{array}{l}
\phi_1\rightarrow e^{-i\alpha/2}\,\phi_1\,,\\
\phi_2\rightarrow e^{i\alpha/2}\,\phi_2\,,\\
\ell_{eL}\rightarrow \ell_{e L}\,,\\
 \ell_{\mu L}\rightarrow
e^{i\alpha}\,\ell_{\mu L}\,,\\
 \ell_{\tau L}\rightarrow e^{-i\alpha}\,\ell_{\tau L}\,,
\end{array}\right.\supset&
\left\{
\begin{array}{l}
\phi_1\rightarrow \phi_1\,,\\
\phi_2\rightarrow \omega_8^2\,\phi_2\,,\\
\ell_{eL}\rightarrow \omega_8^7\,\ell_{e L}\,,\\
 \ell_{\mu L}\rightarrow
\omega_8\,\ell_{\mu L}\,,\\
 \ell_{\tau L}\rightarrow \omega_8^5\,\ell_{\tau L}\,,
\end{array}\right.
\end{array}
\end{equation}
where the minimal discrete Abelian realization $Z_8$ that leads to this
texture can be easily found resorting to the canonical method (cf.
Table~\ref{tablechains} in~\ref{app1}).

The symmetries obtained should not be understood loosely. In the case of
discrete symmetries, these are the minimal symmetries that lead to the
required effective neutrino textures. For continuous symmetries, our
construction implies that, even though there could be some discrete symmetry
implementing such textures, they always lead to a continuous symmetry in the
effective Lagrangian. Notice however that these are symmetries at the
effective level; they are not necessarily present at the UV level. Actually,
depending on the UV completion, we may only need subgroups of the effective
flavour group.

We end this section by analysing the scalar sector. At the effective level we
have only two Higgs doublets, transforming as $\phi_1\rightarrow \phi_1\,,
\phi_2\rightarrow \omega_n\,\phi_2$ under the flavour symmetry. We summarize
in Table~\ref{TabScalar} the minimal discrete groups present in the scalar
potential. Since the scalar potential contains only two Higgs doublets, the
largest discrete Abelian symmetry is $Z_2$~\cite{Ivanov:2011ae}. A larger
group would lead to a continuous accidental $U(1)$. Therefore, in order to
avoid the unwanted (pseudo-) Goldstone bosons, an UV completion of these
models is needed.

In principle, continuous symmetries in the scalar sector are dangerous only
when they are spontaneously broken. If the 2HDM scalar potential produces an
inert-like vacuum expectation value (vev) alignment $\langle\phi_1\rangle
\neq 0,\, \langle\phi_2\rangle = 0$, then the symmetry remains unbroken. But,
in this case, some terms in the texture decomposition will not contribute to
the neutrino mass matrix and the desired texture cannot be constructed. This
argument validates the point that we should avoid scalar $U(1)$'s.

\begin{table}[t]
\caption{\label{TabScalar} Minimal symmetry groups for the 2HDM effective scalar sector.}
\begin{center}
\begin{tabular}{ccc}
\hline
\multirow{2}{*}{Texture}&Symmetry&Symmetry\\
&of the fields&of the potential\\
\hline\\
$\widetilde{\mathbf{A}}$&$Z_5$&\multirow{7}{*}{$U(1)$}\\
\\
\multirow{3}{*}{$\widetilde{\mathbf{B}}$}&$Z_5$&\\
&$Z_4$&\\
&$Z_2\times Z_3$&\\
\\
$\widetilde{\mathbf{C}}$&$Z_4$&\\\\
\hline
\end{tabular}

\end{center}
\end{table}

\section{Symmetry realization of the FGM two-zero textures in a seesaw framework}

Perhaps the simplest UV completions of the effective models previously
discussed are those based on the seesaw models for the neutrino masses. Next
we present the implementations of the minimal Abelian symmetries for the
two-zero neutrino textures in the context of the type I and type II seesaw
mechanisms.

\subsection{Type-II seesaw realization}

In the type II seesaw framework, $SU(2)_L$ triplet scalars $\Delta_k$ with
hypercharge $Y=1$,
\begin{equation}
\Delta_k =
\begin{pmatrix}
\Delta_k^0&-\dfrac{\Delta_k^+}{\sqrt{2}}\\
-\dfrac{\Delta_k^+}{\sqrt{2}}&\Delta_k^{++}
\end{pmatrix}\,,
\end{equation}
are added to the SM particle content.

For our discussion, the relevant terms in the UV completion are
\begin{equation}\label{typeII}
-\mathcal{L}^{\text{II}}_{\text{int}}=
Y^{k}_{\alpha\beta}\overline{\ell_{L\alpha}}\Delta^\dagger_k\ell_{L\beta}^c
+\mu_{k,ab}\tilde{\phi}_a^T\Delta_k\tilde{\phi}_b+ \cdots\,,
\end{equation}
which leads, after the decoupling of the heavy states of mass $M_k$, to the
effective coupling
\begin{equation}
\frac{K^{ab}_{\alpha\beta}}{\Lambda}=-\frac{2\mu_{k,ab}}{M_k^2}Y^k_{\alpha\beta}\,.
\end{equation}
An equivalent way is saying that the Higgs triplets acquire small vevs of the
form
\begin{equation}
\langle\Delta_k^\dagger\rangle=-\frac{\mu_{k,ab}}{M_k^2}
\langle\tilde{\phi}_a\rangle\langle\tilde{\phi}_b\rangle^T.
\end{equation}

In order to extend the analysis previously done for the effective operator,
we just need to replace the field combination $\tilde{\phi}\,\tilde{\phi}^T$
by $\Delta^\dagger$. Thus, we require the flavour charge $\mathcal{Q}_F$,
associated with the flavour group, to satisfy the relation
$\mathcal{Q}_F(\Delta_k)=\mathcal{Q}_F(\phi_i)\,\mathcal{Q}_F(\phi_j)$, and
the corresponding field transformation $\Delta_k\rightarrow
\mathcal{Q}_F(\Delta_k)\Delta_k$. For each different $i$ and $j$ combination
at the effective level, a triplet scalar $\Delta_k$ should be introduced.
This means that effective models with the $\tilde{\phi}_1\tilde{\phi}_2^T$
combination require three Higgs triplets to be implemented. These UV models
will then contain an extended scalar sector with two Higgs doublets and three
Higgs triplets (or two triplets in case (d) of the class
$\widetilde{\mathbf{B}}$ given in Eq.~\eqref{decompclassB}). The field
transformations are
\begin{equation}
\begin{array}{l}
\phi_1\rightarrow \phi_1\,,\quad \phi_2\rightarrow\omega_n\,\phi_2\,,\\
\Delta_1\rightarrow \Delta_1\,,\quad \Delta_2\rightarrow \omega_n^2\,\Delta_2\,,
\quad \Delta_3 \rightarrow \omega_n\,\Delta_3\,.
\end{array}
\end{equation}

Since we have considerably enlarged the scalar sector, it is important to
check whether accidental symmetries can be avoided in the scalar potential.
For a model with two scalar doublets and several scalar triplets, the phase
sensitive terms in the scalar potential are $\phi_a^\dagger\,\phi_b$,
$(\phi_a^\dagger\,\phi_b)^2$, $\Delta_i^\dagger\,\Delta_j$,
$(\Delta_i^\dagger\,\Delta_j)^2$, $\Delta_i^\dagger\,\Delta_j\,
\Delta_k^\dagger\,\Delta_l$, $\phi_a^\ast\, \phi_b^\ast\, \Delta_k$, and
$\phi^\dagger_a\, \phi_b\, \Delta_i^\dagger\, \Delta_j$.

A $Z_n$ group is a symmetry of the scalar potential if it does not induce a
larger symmetry. Therefore, one needs to check for terms that under the field
transformations transform with $\omega_n^k$. For example, the term
$\phi_a^\ast\, \phi_b^\ast\, \Delta_k$, even though is phase sensitive, it is
by construction invariant and, consequently, not sensitive to the order of
the group. On the other hand, the term $\phi_1^\dagger \,\phi_2\,
\Delta_1^\dagger\, \Delta_2$ transforms with $\omega_n^3$. The presence of
this term in the potential implies $n=3$, that is, the group is $Z_3$ and not
$U(1)$. It is easy to check that the term with the largest phase
transformation is $(\Delta_1^\dagger \, \Delta_2)^2$, which transforms with
$\omega_n^4$. Therefore, the largest discrete Abelian symmetry allowed in a
2HDM plus three (or two) scalar triplets is $Z_4$. From
Table~\ref{TabScalar}, we then conclude that cases leading to a $Z_5$
symmetry will have an accidental continuous symmetry in the scalar potential.
Other cases, including the $Z_2\times Z_3$ one, can have an UV completion in
a type II seesaw framework without introducing continuous symmetries in the
scalar potential.

Defining the field vector $(\phi_a,\,\Delta_k)$ we can determine the $D$
matrix for each case. Consider first the $Z_8$ case in class
$\widetilde{\mathbf{B}}$. Since the scalar phase transformation of $\phi_2$
is $\omega_8^2=\omega_4$, the scalar field transformations are given by
\begin{equation}
\begin{array}{l}
\phi_1\rightarrow \phi_1\,,\quad \phi_2\rightarrow \omega_4\,\phi_2\,,\\
\Delta_1\rightarrow \Delta_1\,,\quad \Delta_2\rightarrow \omega_4^2\,\Delta_2\,,
\quad\Delta_3\rightarrow \omega_4\,\Delta_3\,.
\end{array}
\end{equation}
The set of phase sensitive interactions yields the matrix
\begin{equation}\label{DBZ8}
D=
\begin{pmatrix}
-2&0&1&0&0\\
0&-2&0&1&0\\
-1&-1&0&0&1\\
-1&1&1&0&-1\\
-1&1&0&-1&1\\
0&0&-2&2&0
\end{pmatrix},
\end{equation}
which leads to $D_{SNF}=\text{diag}(1,1,1,4,0)$, as expected. Therefore, this
model has a $Z_8$ flavour symmetry in the full (scalar + leptonic)
Lagrangian. Nevertheless, in the scalar potential only the $Z_4$ subgroup
acts nontrivially.

Next let us consider the $Z_2\times Z_2\times Z_3$ case of class
$\widetilde{\mathbf{B}}$. The scalar field transformations are now given by
\begin{equation}
\begin{array}{l}
\phi_1\rightarrow (1,1,1)\,\phi_1\,,\quad \phi_2\rightarrow (1,-1,\omega)\,\phi_2\\
\Delta_1\rightarrow (1,1,1)\,\Delta_1\,,\quad \Delta_2\rightarrow (1,1,\omega^2)\,\Delta_2\,.
\end{array}
\end{equation}

The first $Z_2$ is irrelevant, since it simply reflects a discrete lepton
number. Therefore, we focus on the possible implementation of $Z_2\times Z_3$
in the scalar potential with two Higgs doublets and two Higgs triplets. As
already pointed out, this implies an accidental symmetry in the scalar
potential. There is, however, a way out in this case. Looking at the symmetry
implementation in the effective operator, one sees that while $Z_3$ gives the
flavour structure, the only purpose of $Z_2$ is to forbid the cross term
$\phi_1^\ast\,\phi_2^\ast$, i.e. it is just a shaping symmetry. In our type
II implementation, the above cross term has been avoided by removing the
triplet scalar $\Delta_3$ associated with it, and keeping only
$\Delta_{1,2}$. Therefore, at the UV level we only have the $Z_3$ symmetry
for the scalar fields, implying
\begin{equation}
D=
\begin{pmatrix}
-2&0&1&0\\
0&-2&0&1\\
-1&1&-1&1
\end{pmatrix},
\end{equation}
with the Smith normal form $D_{SNF}=\text{diag}(1,1,3,0)$, as expected. The
full Lagrangian has a $Z_2\times Z_3$ symmetry, where the $Z_2$ corresponds
to the discrete lepton number. After the decoupling of the heavy triplets,
the effective Lagrangian sees its symmetry being enlarged.

Finally we consider class $\widetilde{\mathbf{C}}$. Although the flavour
symmetry is continuous in the effective approach, it may be accidental. As
can be seen from Eq.~\eqref{CaseC}, the effective texture can be implemented
by a $Z_8$ symmetry. In the scalar sector, the field transformations are
exactly the same as in case $Z_8$ of class $\widetilde{\mathbf{B}}$.
Therefore, for the scalar sector, $D$ has the same form of Eq.~\eqref{DBZ8}.
This implies that no accidental symmetries appear in the scalar sector. The
model can then be implemented from a $Z_8$ flavour symmetry group. The Yukawa
sector alone will still have an accidental global $U(1)$ symmetry, as can be
checked by constructing the $D$ matrix for that sector only. Yet the full
Lagrangian and, most importantly, the scalar potential has only a discrete
symmetry avoiding unwanted (pseudo-) Goldstone bosons.

We summarize our results in Table~\ref{TabTypeII}, where we present the
allowed symmetries in the effective 2HDM and in the corresponding type II
seesaw UV completion.

\begin{table}[t]
\caption{\label{TabTypeII} Symmetries in the effective model and type II UV completion.
The trivial $Z_2$ associated with lepton number has been omitted.}
\begin{center}
\begin{tabular}{ccccc}
\hline
\multirow{2}{*}{Texture}&Effective&2HDM\,+&Scalar&Goldstone\\
&2HDM&type II&potential&boson\\
\hline
&&&&\\
$\widetilde{\mathbf{A}}$&$U(1)$&$U(1)$&$U(1)$&Yes\\
&&&&\\
\hline
&&&&\\
\multirow{3}{*}{$\widetilde{\mathbf{B}}$}&$ Z_5$&$ Z_5$&$U(1)$&Pseudo\\
&$Z_8$&$Z_8$&$Z_4$&No\\
&$ Z_2\times Z_3$&$Z_3$&$Z_3$&No\\
&&&&\\
\hline
&&&&\\
$\widetilde{\mathbf{C}}$&$U(1)$&$Z_8$&$Z_4$&No\\
&&&&\\
\hline
\end{tabular}
\end{center}
\end{table}

Some of the symmetry implementations presented in this section have been
previously studied~\cite{Grimus:2004az,Fritzsch:2011qv}. However, the
discrete symmetry groups differ in some of them. In
Refs.~\cite{Grimus:2004az,Fritzsch:2011qv}, textures $\mathbf{A}_{1,2}$ and
$\mathbf{B}_{3,4}$, which in our case belong to class
$\widetilde{\mathbf{A}}$, are implemented with $Z_6$. As we have shown, any
discrete group $Z_k$, with $k \geq 5$, implies a global accidental $U(1)$ in
the full Lagrangian. The textures $\mathbf{B}_{1,2}$ are implemented in the
above works using two Higgs triplets and a $Z_3$ symmetry. This is precisely
our last case of class $\widetilde{\mathbf{B}}$. There are, however, two
other minimal implementations of these zero textures, which are given in
Table~\ref{TabTypeII}. Concerning the implementation of texture $\mathbf{C}$,
in the above works a $Z_4$ discrete symmetry is used, but the scalar Higgs
doublets were ignored. As shown before, while $Z_4$ is the symmetry of the
scalar potential, the full Lagrangian requires a $Z_8$ flavour group. We also
note that in all the implementations of
Refs.~\cite{Grimus:2004az,Fritzsch:2011qv} the flavour symmetry needs to be
softly broken in the scalar potential. This is due to the fact that only one
Higgs doublet is introduced in the theory. It is precisely the existence of
such soft-breaking terms that permits the implementation of $\mathbf{C}$ with
a $Z_4$ and not a $Z_8$ symmetry.

A final remark on the Type-II UV completions is in order. We have enlarged
the scalar sector in such a way that, in some cases, it is enough to have no
accidental continuous symmetries in the scalar potential. These extensions
may solve the global $U(1)$ scalar problem, but they may have difficulties
with the presence of very light scalar particles. In order to see this, let
us take as an example the $Z_8$ case in class $\widetilde{\mathbf{B}}$. There
is a single term that requires a $Z_4$ symmetry in the scalar potential,
namely, $\lambda_\Delta(\Delta^\dagger_1\Delta_2)^2$. In the limit
$\lambda_\Delta \rightarrow 0$, we recover the $U(1)$ symmetry of the
effective potential. This term appears in the scalar mass matrix
diagonalization, once $\phi_a$ and $\Delta_i$ acquire vevs. However,
$\langle\Delta_i\rangle$ should be very small in order to explain the
neutrino masses. This implies that the corrections to the massless scalars
are also very small.

\subsection{Type-I seesaw realization}

In the canonical type I seesaw scenario, three right-handed neutrinos,
$N_{iR}$, with heavy masses in order to explain the light neutrino masses,
are added to the SM particle content. The UV Lagrangian is of the form
\begin{equation}\label{typeI}
-\mathcal{L}^{\text{I}}_{\text{int}}=
Y^{a}_{\alpha i}\overline{\ell_{L\alpha}}\tilde{\phi}_aN_{iR}
+\frac{1}{2}M_{ij}\overline{N_{iR}^c}N_{jR}\,,
\end{equation}
which leads, after the decoupling of the heavy states, to the effective
coupling
\begin{equation}
\frac{K^{ab}_{\alpha\beta}}{\Lambda}=-Y^aM^{-1}Y^{bT}\,.
\end{equation}
Contrarily to the type II seesaw, in this UV completion the effective
coupling is not directly extracted from the UV Lagrangian, which usually
makes the construction of these models more challenging. In order to obtain
the possible implementations, we recall that the generator $\mathcal{S}_L$
has no degeneracy in all possible effective implementations. It remains to
find the way that the heavy right-handed neutrino fields transform under the
flavour symmetry. Up to permutations, we can split the analysis into three
cases: $\mathcal{S}_R=\pm\,\mathbb{1}$, $\mathcal{S}_R=\pm\,
\text{diag}\,(1,1,-1)$, and $\mathcal{S}_R=\text{diag}\,(\pm
1,e^{i\beta},e^{-i\beta})$ with $\beta \neq \pi$.

In the first case, the matrix $M$ is completely general (texture $A_1$), and
the Yukawa textures contain just a line of nonzero entries. One can easily
check that $Y^a M^{-1} Y^{aT} \sim Y^aY^{aT}$ has a diagonal texture (some
diagonal entries may be zero). Since none of the realizable cases has two
textures of the diagonal form, this scenario is excluded.

In the second case, the matrix $M$ (and its inverse) has a block-diagonal
form (texture $A_2$). Once again, it turns out that $Y^aM^{-1}Y^{aT} \sim Y^a
Y^{aT}$ has a diagonal texture. Therefore, this case is also excluded.

There remains the case with $\mathcal{S}_R=\text{diag}(\pm
1,e^{i\beta},e^{-i\beta})$. In this case, the right-handed neutrino mass
matrix is given by
\begin{equation}\label{MR}
M=
\begin{pmatrix}
\times&&\\
&&\times\\
&\times&
\end{pmatrix}=P_{23} A_{13}.
\end{equation}

Since all the phase transformations are known, each possibility can be
straightforwardly analysed. For class $\widetilde{\mathbf{A}}$, the only
possible implementation up to permutations on the right side is given by
Eq.~\eqref{ClassA}, with the additional field transformations
\begin{equation}
Z_2 \times U(1):\left\{
\begin{array}{l}
\begin{array}{l}
N_1\rightarrow (-1,1)N_1,\quad N_2\rightarrow (1,e^{i\alpha})N_2,\\
N_3\rightarrow (1,e^{-i\alpha})N_3\,.
\end{array}
\end{array}\right.
\end{equation}
Note that $U(1)$ is the group that completely defines the flavour structure
of the couplings, while the additional $Z_2$ reflects an accidental symmetry
when that structure is present. At the UV level, the fields $\ell_{\alpha L}$
and $\phi_a$ will transform exactly as in Eq.~\eqref{ClassA} under the $U(1)$
group. However, under the $Z_2$ symmetry their transformation is now given by
\begin{equation}
Z_2:\quad\ell_{\mu L}\rightarrow -\ell_{\mu L}\,,\quad \phi_2\rightarrow -\phi_2\,,
\end{equation}
with the remaining fields invariant. The Yukawa couplings are given by
\begin{equation}
Y^\nu_1=
\begin{pmatrix}
&\times&\\
\times&&\\
&&\times
\end{pmatrix}\,,\quad
Y^\nu_2=
\begin{pmatrix}
&&\\
&\times&\\
\times&&
\end{pmatrix}\,.
\end{equation}
The minimal discrete implementation is then obtained with the replacement
$e^{i\alpha}\rightarrow \omega_5$.

It is instructive to make the connection between the $Z_2$ symmetry at high
and low energies. Looking at the way $\ell_{\alpha L}$ and $\phi_a$ transform
at high energies, it is not evident that this symmetry corresponds to the
discrete lepton number at low energies. To check this, let us decouple the
right-handed fields and write the effective Lagrangian as
\begin{align}
\begin{split}
-\mathcal{L}_{\text{int}} \sim&
\left(\overline{\ell_{eL}}\,\ell_{\tau L}^c+\overline{\ell_{\tau L}}\,\ell_{eL}^c
+\overline{\ell_{\mu L}}\,\ell_{\mu L}^c\right)
\tilde{\phi}_1\tilde{\phi}_1^T\\
&+
\left(\overline{\ell_{\mu L}}\,\ell_{\tau L}^c+
\overline{\ell_{\tau L}}\,\ell_{\mu L}^c\right)
\tilde{\phi}_1\tilde{\phi}_2^T+
\overline{\ell_{\tau L}}\,\ell_{\tau L}^c
\tilde{\phi}_2\tilde{\phi}_2^T\,,
\end{split}
\end{align}
where, for simplicity, we have omitted the coupling in each term. The terms
of type $\phi_a^2$ are insensitive to the change $\ell_{\mu L}\rightarrow
-\ell_{\mu L}$ under $Z_2$. In the cross term, this field transformation is
indistinguishable from demanding $\phi_1 \rightarrow -\phi_1$. The question
is whether the latter replacement leads to any change in the other Lagrangian
terms. If $\phi_1$ couples to the charged lepton Yukawa term, then the
$e_{\beta R}$ charges can be properly chosen to account for this
transformation. Concerning the scalar potential, since it has a $U(1)$ global
symmetry the only cross term is $|\phi^\dagger_1\phi_2|^2$, which is
insensitive to the above transformation. Therefore, at low energies the $Z_2$
symmetry can be expressed as in Eq.~\eqref{ClassA}.

In the case of class $\widetilde{\mathbf{B}}$, there are only discrete
groups. Checking all possible charge assignments to $N_{iR}$ under the
flavour group, we found no viable implementation within this class.

Finally, for class $\widetilde{\mathbf{C}}$ the only possible implementation
up to permutations on the right is given by Eq.~\eqref{CaseC} with the
additional field transformations
\begin{equation}
 Z_2\times U(1):\left\{
\begin{array}{l}
\begin{array}{l}
N_1\rightarrow (-1,1)N_1,\quad N_2\rightarrow (1,e^{i\alpha/2}) N_2,\\
N_3\rightarrow (1,e^{-i\alpha/2})N_3\,.
\end{array}
\end{array}\right.
\end{equation}
The Yukawa textures take the form
\begin{equation}
Y^\nu_1=
\begin{pmatrix}
&&\times\\
&\times&\\
&&
\end{pmatrix}\,,\quad
Y^\nu_2=
\begin{pmatrix}
&\times&\\
&&\\
&&\times
\end{pmatrix}\,.
\end{equation}
The discrete implementation is obtained with the replacement
$e^{i\alpha/2}\rightarrow \omega_8$.

The results for all cases are summarized in Table~\ref{TabTypeI}, where the
allowed symmetries in the effective 2HDM and the corresponding type I seesaw
UV completions are given.\footnote{A symmetry realization of the
phenomenologically viable FGM two-zero texture neutrino mass matrices within
the framework of the mixed type-I + type-II seesaw mechanism has been
considered in Ref.~\cite{Dev:2011jc}.}

\begin{table}[h]
\caption{\label{TabTypeI} Symmetries in the effective model and type I UV completion.
The trivial $Z_2$ associated with lepton number has been omitted.}
\begin{center}
\begin{tabular}{ccc}
\hline
Texture&Effective 2HDM & 2HDM + type I\\
\hline
&&\\
$\widetilde{\mathbf{A}}$&$U(1)$&$Z_2\times U(1)$\\
&&\\
\hline
&&\\
\multirow{3}{*}{$\widetilde{\mathbf{B}}$}&$ Z_5$&\\
&$Z_8$&Non-realizable\\
&$ Z_2\times Z_3$&\\
&&\\
\hline
&&\\
$\widetilde{\mathbf{C}}$&$U(1)$&$Z_2\times U(1)$\\
&&\\
\hline
\end{tabular}
\end{center}
\end{table}

\section{Conclusions}

We have obtained the minimal Abelian symmetry realizations of
phenomenological two-zero neutrino textures, i.e. neutrino mass matrices with
two zeros, written in the physical basis where the charged leptons are
diagonal. The symmetry constructions were achieved resorting to the canonical
and Smith normal form methods. The implementation of these symmetries in UV
completions based on the type I and type II seesaw mechanisms was also
presented. It is worth noticing that the discrete symmetry realizations of
the two-zero neutrino textures presented here are different from previous
studies~\cite{Grimus:2004hf,Grimus:2004az,Fritzsch:2011qv}. Indeed, in our
implementations the flavour symmetry in the leptonic sector is only broken at
the electroweak scale and not at the (high) seesaw scale. This means, in
particular, that the texture zeros in the effective neutrino mass matrix will
remain exact up to the electroweak scale, without being affected by
renormalization group corrections.

Finally, we also remark that the minimal effective and seesaw-like
implementations of the neutrino textures typically suffer from the presence
of very light (or even massless) scalars. This is due to the existence of
only two Higgs doublets at the electroweak scale, transforming under an
Abelian group of order greater than two. In this work, we have focused on the
leptonic sector only; it may happen that quarks interact with other scalar
doublets or that additional scalar doublets, associated with repeated
textures, appear in the neutrino sector. In such cases, larger symmetries of
the scalar potential are allowed. Independently of the implementation chosen,
if we insist on curing this problem without breaking softly the flavour
symmetry, we need to extend the 2HDM scalar potential at the electroweak
scale.

\section*{Acknowledgements}

We are grateful to  I.P.~Ivanov for enlightening comments and suggestions.
The work of R.G.F. was partially supported by FCT - \textit{Funda\c{c}\~{a}o para a
Ci\^{e}ncia e a Tecnologia}, under the projects PEst-OE/FIS/UI0777/2013 and
CERN/FP/123580/2011. The work of H.S. is funded by the European FEDER,
Spanish MINECO, under the grant FPA2011-23596, and the Portuguese FCT project
PTDC/FIS-NUC/0548/2012.

\appendix

\section{Discrete symmetries, texture decompositions and charges}
\label{app1} \setcounter{table}{0} \renewcommand{\thetable}{A.\arabic{table}}

\begin{table*}[t]
\caption{Discrete symmetry, two-zero texture decomposition and  Higgs
combination charges obtained from the canonical method.
Here $P=\left\{\mathbb{1},P_{12},P_{13},P_{23},P_{123},P_{321}\right\}$,
$P^\prime=\left\{\mathbb{1},P_{12},P_{13}\right\}$ and
$P^{\prime\prime}=\left\{\mathbb{1},P_{13},P_{23}\right\}$.}
\label{tablechains}
\begin{center}
\begin{tabular}{rll}
\hline
Symmetry&Texture decomposition& Charges\\
\hline\\
None&$A_1$\\
&\\
$Z_{2n}$&$P\left\{A_2\oplus A_3\right\}P$
&$\left(1,\,\omega_{2n}^n\right); \, k=n$\\
&\\
$Z_{n\geq 3}$&$P\left\{A_7\oplus A_3\oplus A_{12}\right\}P$
&$\left(1,\,\omega_n^{k},\, \omega_n^{2k}\right)$\\
&\\
$Z_{3n}$&$A_{13}P_{23}\oplus A_{13}P_{12}\oplus A_{13}P_{13}$
&$\left(1,\, \omega_{3n}^n,\,\omega_{3n}^{2n}\right); \, k_1=-k_2=n$\\
&\\
$Z_{4n}$&$P^\prime\left\{A_{13}P_{23}\oplus P_{13}A_{15}P_{123}\oplus A_{15}
\oplus P_{123}A_{15}P_{12}\right\}P^\prime$
&$\left(1,\, \omega_{4n}^n,\, \omega_{4n}^{2n},\, \omega_{4n}^{3n}\right); \, k_1=-k_2=n$\\
&\\
$Z_{n\geq 5}$&$P^\prime\left\{A_{13}P_{23}\oplus P_{13}A_{15}P_{123}
\oplus P_{123}A_{15}P_{12}\right.$\\
&$\left.\oplus P_{23}A_{12}P_{23}\oplus A_{12}\right\}P^\prime$
&$\left(1,\,\omega_n^{k_1},\,\omega_n^{-k_1},\,\omega_n^{2k_1},\,
\omega_n^{-2k_1}\right); \, k_2=-k_1$\\
&\\
$Z_{2(n+2)}$&$P^{\prime\prime}\left\{P_{13}A_{15}P_{13}\oplus P_{13}A_{15}P_{123}
\oplus P_{123}A_{15}P_{12}\right.$\\
&$\left.\oplus A_{15}P_{23}\oplus A_{12}\oplus A_0\right\}P^{\prime\prime}$
&$\left(1,\,\omega_{2(n+2)}^{n+2},\,\omega_{2(n+2)}^{k_{2}},\,
\omega_{2(n+2)}^{k_{2}+n+2},\,\omega_{2(n+2)}^{2k_{2}}\right); \, k_1=n+2$\\
&\\
$Z_{n\geq 6}$&$P_{13}A_{12}P_{13}\oplus P_{13}A_{15}P_{123}
\oplus P_{123}A_{15}P_{12}\oplus A_{15}P_{23}$\\
&$\oplus P_{23}A_{12}P_{23}\oplus A_{12}$
&$\left(1,\,\omega_n^{k_{1}},\,\omega_n^{k_{2}},\,\omega_n^{k_{2}+k_{1}},\,
\omega_n^{2k_{1}},\,\omega_n^{2k_{2}}\right)$\\
&\\
$Z_{2n}\times Z_{2m}$&$P \left\{ A_{13}\oplus P_{321}A_{15}P_{13}\oplus
P_{123}A_{15}P_{12}\oplus A_{15}P_{23}\right\}P$
&$\left[(1,1),(1,\omega_{2m}^m),(\omega_{2n}^n,1),(\omega_{2n}^n,\omega_{2m}^m)
\right];\, k=n,\, k^\prime=m$\\\\
\hline
\end{tabular}
\end{center}
\end{table*}

In this appendix we construct the minimal discrete implementations of the
allowed textures, and the corresponding phase transformation matrices, using
the canonical method~\cite{Serodio:2013gka}. In order to exemplify the
procedure we consider the simple case when
$\mathcal{S}_L=\text{diag}(1,1,e^{i\alpha})$. We start by replacing the
continuous phase by a discrete one, i.e. $e^{i\alpha}\rightarrow \omega_n^k$,
where the index $n$ is the order of the group and $k$ is the integer
associated charge. The phase transformation matrix $\Theta$ is a matrix
containing in its entries the associated phase transformation of the lepton
fields. Therefore, for a term of the form $L^\ast L^\ast \Sigma$, with
$\Sigma$ a scalar field, we get
\begin{equation}
\Theta=
\begin{pmatrix}
1&1&\omega_n^{-k}\\
1&1&\omega_n^{-k}\\
\omega_n^{-k}&\omega_n^{-k}&\omega_n^{-2k}
\end{pmatrix}\,.
\end{equation}
From this matrix we obtain two different implementations: $\omega_n^{-2k}=1$
or $\omega_n^{-2k}\neq 1$.

If $\omega_n^{-2k}=1$, then $2k=0\, (\text{mod}\,n)$ and the order of the
group has to be a multiple of two. We define the group as $Z_{2n}$. The
implementation, or chain, is then given by $A_2 \oplus A_3$ with the
associated charges $(1,\omega_{2n}^n)$. When $\omega_n^{-2k}\neq 1$, $k\neq
2k\neq 0$ and the order of the group has to be $n \geq 3$. We define the
group as $Z_{n\geq 3}$. The corresponding chain is given by $A_7\oplus
A_3\oplus A_{12}$, with the charges $\left(1,\,\omega_n^{k},\,
\omega_n^{2k}\right)$.

The above procedure can be repeated for the remaining cases. For the cases
where there is no degeneracy, the discrete form of the left generator is
given by $\mathcal{S}_L=\text{diag}(1,\omega_n^{k_1},\omega_n^{k_2})$. The
results are displayed in Table~\ref{tablechains}.

Note that the vector of charges indicates the way that the auxiliary field
$\Sigma$ transforms in order to pick a given texture. For instance, in the
type II seesaw, the field $\Sigma$ is identified with the scalar triplet
$\Delta^\dagger$, while in the general dimension-5 effective approach it is
identified with the Higgs doublet combination $\phi_a^\ast \phi_b^\ast$.

\end{document}